\documentstyle{mn}

\newif\ifAMStwofonts

\def\msun{{\rm\,M_\odot}}
\def\msun{{\rm\,M_\odot}}

\def\spose#1{\hbox to 0pt{#1\hss}}
\def\lta{\mathrel{\spose{\lower 3pt\hbox{$\mathchar"218$}}
     \raise 2.0pt\hbox{$\mathchar"13C$}}}
\def\gta{\mathrel{\spose{\lower 3pt\hbox{$\mathchar"218$}}
     \raise 2.0pt\hbox{$\mathchar"13E$}}}

\def\bx{{\bf x}}
\def\bI{{\bf I}}
\def\barI{{\bar I}}

\def\bxi{{\bf\xi}}
\def\bOmega{{\Omega}}           
\def\nat{\rm Nature}
\def\apj{\rm ApJ}
\def\aj{\rm AJ}
\def\mnras{\rm MNRAS}

\ifoldfss
  \ifCUPmtlplainloaded \else
    \NewTextAlphabet{textbfit} {cmbxti10} {}
    \NewTextAlphabet{textbfss} {cmssbx10} {}
    \NewMathAlphabet{mathbfit} {cmbxti10} {} 
    \NewMathAlphabet{mathbfss} {cmssbx10} {} 
  \fi
  \ifAMStwofonts
    \ifCUPmtlplainloaded \else
      \NewSymbolFont{upmath} {eurm10}
      \NewSymbolFont{AMSa} {msam10}
      \NewMathSymbol{\upi}     {0}{upmath}{19}
      \NewMathSymbol{\umu}     {0}{upmath}{16}
      \NewMathSymbol{\upartial}{0}{upmath}{40}
      \NewMathSymbol{\leqslant}{3}{AMSa}{36}
      \NewMathSymbol{\geqslant}{3}{AMSa}{3E}

       \let\ge=\geqslant
    \fi
  \fi
\fi 

\ifnfssone
  \newmathalphabet{\mathit}
  \addtoversion{normal}{\mathit}{cmr}{m}{it}
  \addtoversion{bold}{\mathit}{cmr}{bx}{it}
  \newmathalphabet{\mathbfit} 
  \addtoversion{normal}{\mathbfit}{cmr}{bx}{it}
  \addtoversion{bold}{\mathbfit}{cmr}{bx}{it}
  \newmathalphabet{\mathbfss} 
  \addtoversion{normal}{\mathbfss}{cmss}{bx}{n}
  \addtoversion{bold}{\mathbfss}{cmss}{bx}{n}
  \ifAMStwofonts
    \ifCUPmtlplainloaded \else
      \UseAMStwoboldmath
      \makeatletter
      \new@mathgroup\upmath@group
      \define@mathgroup\mv@normal\upmath@group{eur}{m}{n}
      \define@mathgroup\mv@bold\upmath@group{eur}{b}{n}
      \edef\UPM{\hexnumber\upmath@group}
      \new@mathgroup\amsa@group
      \define@mathgroup\mv@normal\amsa@group{msa}{m}{n}
      \define@mathgroup\mv@bold\amsa@group{msa}{m}{n}
      \edef\AMSa{\hexnumber\amsa@group}
      \makeatother
      \mathchardef\upi="0\UPM19
      \mathchardef\umu="0\UPM16
      \mathchardef\upartial="0\UPM40
      \mathchardef\leqslant="3\AMSa36
      \mathchardef\geqslant="3\AMSa3E

       \let\ge=\geqslant
    \fi
  \fi
\fi 

\ifnfsstwo
  \DeclareMathAlphabet{\mathbfit}{OT1}{cmr}{bx}{it}
  \SetMathAlphabet\mathbfit{bold}{OT1}{cmr}{bx}{it}
  \DeclareMathAlphabet{\mathbfss}{OT1}{cmss}{bx}{n}
  \SetMathAlphabet\mathbfss{bold}{OT1}{cmss}{bx}{n}
  \ifAMStwofonts
    \ifCUPmtlplainloaded \else
      \DeclareSymbolFont{UPM}{U}{eur}{m}{n}
      \SetSymbolFont{UPM}{bold}{U}{eur}{b}{n}
      \DeclareSymbolFont{AMSa}{U}{msa}{m}{n}
      \DeclareMathSymbol{\upi}{0}{UPM}{"19}
      \DeclareMathSymbol{\umu}{0}{UPM}{"16}
      \DeclareMathSymbol{\upartial}{0}{UPM}{"40}
      \DeclareMathSymbol{\leqslant}{3}{AMSa}{"36}
      \DeclareMathSymbol{\geqslant}{3}{AMSa}{"3E}

       \let\ge=\geqslant
    \fi
  \fi
\fi 

\ifCUPmtlplainloaded \else
  \ifAMStwofonts \else 
    \def\upi{\pi}
    \def\umu{\mu}
    \def\upartial{\partial}
  \fi
\fi

\begin{document}

\onecolumn

\title{Noise-driven evolution in stellar systems: Theory}
\author[Martin D. Weinberg]
{Martin D. Weinberg \\
Department of Astronomy, University of Massachusetts, Amherst, MA 01003-4525, USA}

\date{}
\pagerange{\pageref{firstpage}--\pageref{lastpage}}
\pubyear{}

\maketitle

\label{firstpage}

\begin{abstract}
  We present a theory for describing the evolution of a galaxy caused
  by stochastic events such as weak mergers, transient spiral
  structure, orbiting blobs, etc.  This noise excites large-scale
  patterns that drives the evolution of the galactic density profile.
  In dark-matter haloes, the repeated stochastic perturbations
  preferentially ring the lowest-order modes of the halo with only a
  very weak dependence on the details of their source.  Shaped by
  these modes, the profile quickly takes on a nearly self-similar
  form.  We show that this form has the features of the ``universal
  profile'' reported by Navarro, Frenk, \& White independent of
  initial conditions in a companion paper.  In this sense, this
  noise-driven process is a near-equilibrium form of violent
  relaxation.
\end{abstract}

\begin{keywords}
galaxies:evolution --- galaxies: haloes --- galaxies: kinematics and
dynamics --- cosmology: theory --- dark matter
\end{keywords}

\section{Introduction}

Both numerical experiments and observations suggest a universality in
galaxy profiles resulting from dissipationless collapse.  Remnants
from merger simulations (e.g. van Albada 1982, Barnes
1989) typically manifest the often
observed $r^{1/4}$-type profile in elliptical galaxies.  More
recently, the same feature has arisen in the context of cosmological
large-scale structure simulations.  Researchers beginning with
Navarro, Frenk \& White (1997) have noted that CDM
collapse results in a haloes with a similar overall profile.  Debate
remains about the precise functional form of these profiles, but these
and the $r^{1/4}$-law are all quite similar, suggesting that some
general violent-relaxation-like dynamical mechanism is at work.

A precise explanation of the dynamics of violent relaxation, the
process driving the convergent evolution in by large fluctuations in
potential, remains elusive.  There have been several attacks on this
problem.  Tremaine, Henon \& Lynden-Bell
(1986) follow a statistical mechanics
approach and explore extremizing functionals (see their paper for a
review of prior work).  Spergel \& Hernquist
(1992) extended this approach with the
additional ansatz that the orbit--perturbation interaction can be
approximated by kicks near perigalacticon.  Recently, this has been
followed by Mangalam, Nityananda \& Sridhar
(1999) who incorporate these physical
mechanisms in a semi-analytic evolution equation based on diffusion in
action space.  Over the same period, Kandrup has pursued a
mathematically rigorous exploration into the geometry of phase space
resulting from the details of Hamiltonian flows (e.g. see Kandrup 1998
and references therein).

In this paper, I take a different approach to the problem.  Rather
than trying to identify an appropriate form for the phase-space
distribution to start, I treat the problem as an initial value problem
for a stochastic process and develop an evolution equation for
exploring its consequence on stellar systems.  A companion paper
(Paper 2) will apply this to the evolution of galactic haloes in
particular.  The underlying motivation is as follows.  Previous work
by the author has shown that fluctuations in stellar systems on the
largest scales can be strongly amplified by their own self gravity
(e.g. Weinberg 1993).  This means that large-scale
fluctuations will greatly exceed their Poisson amplitudes.  In
particular, Weinberg (1993) explored the idealized case of periodic
cube; the fluctuations become very large as the system size approaches
its Jeans' length.  Similarly for stable galaxies, the fluctuations in
a system will be largest at its discrete modes.  In addition, Weinberg
(1994) argues that galaxies will often have very
weakly damped $m=1$ (sloshing or seich) modes and these result in
large excitations when excited (see Vesperini \& Weinberg 2000 for
another example).  Putting this
together, one might ask: if noise preferentially excites particular
modes independent of the noise source, is it possible the repetitive
stochastic response of the galaxy will lead to some characteristic
features, independent of its initial conditions?

In order to answer this question, this paper concentrates on the
theoretical framework for describing the evolution due to stochastic
fluctuations.  Beginning with a description of the linear response of
a galaxy to excitation, and assuming that the process is Markovian,
one may expand the Boltzmann collision term in a series.  Analogous to
two-body collisions, only the first two terms contribute and the
resulting evolution equation has a Fokker-Planck form.  This equation
is very far from being analytically tractable because the diffusion
coefficients depend on integrals over all of phase space under the
stochastic perturbation.  It does not solve the classic violent
relaxation problem because this approach is perturbative and assumes
that the stellar system remains near an equilibrium, Nonetheless it
does incorporate the same underlying processes, the self-gravitating
response to noise, and yields profiles with many of the same features
that are found in the numerical experiments as Paper 2 will
demonstrate.

The plan for this paper is as follows.  The overall approach is
outlined in \S\ref{sec:overview} followed by an explicit derivation
for spherical haloes in \S\ref{sec:fp}.  This could be
straightforwardly extended to many standard geometries geometry (e.g.
disk or disk and halo together).  We describe the character of the
noise for two general cases, transient and quasi-periodic perturbers
(a halo of black holes, for example) in \S\ref{sec:noise}.  These two
cases result in qualitatively different behavior and represent the
most plausible astronomical scenarios.  Transient noise is probably
the most relevant and important.  Finally, the main features are
summarized in \S\ref{sec:sum}.  Paper 2 will review the basic physics
and apply these methods to understanding the evolution of halo in a
noisy environment, e.g. just after formation, and may be a better
place to begin for those interested in astronomical consequences
rather than the kinetic theory.

\section{Derivation of the evolution equation}
\label{sec:derive}

\subsection{Overview}
\label{sec:overview}

The outer halo in large galaxies is less than 10 dynamical times old
so primordial inhomogeneities will not have had time to phase mix and
continued disturbances from mergers will not have relaxed (see
Tremaine 1992 for additional discussion).  These, as
well as any intrinsic sources of noise, e.g. a population of
$10^6\msun$ massive black holes, dwarf galaxies, debris streams
(Johnston 1998, Morrison et al.
2000) or dark clusters, are
amplified by the self-gravity of the halo.  In the current epoch,
these distortions create potentially {\em observable} asymmetries in
the stellar and gaseous Galactic disk.  Just after galaxy formation,
this noise may be sufficient to drive the evolution of the halo (Paper
2).  Other possible applications include the evolution of proto-stellar
clusters and the evolution of the proto-stellar binary distribution in
the noisy star forming environment (work in progress).  All of these
problems motivated the development of a stellar dynamical kinetic
equation that can handle general stochastic processes.

The general problem will be familiar to most dynamicists and could be
solved following the standard two-body approach: beginning with the
collisional Boltzmann equation, one writes the collision term in
Master equation form and expands in a Taylor series to derive
Fokker-Planck equation (e.g. Binney \& Tremaine 1987, Spitzer
1987).  For a spherically
symmetric system, the phase-space distribution is a function of two
action and and two angle variables.  Averaging over times short
compared to the relaxation time but long compared to the dynamical
time (orbit-averaging), one obtains a Fokker-Planck equation.

Alternatively, recent work in statistical mechanics and noise theory
in general has developed a body of methods for treating stochastic
differential equations based directly on transition probabilities.  If
$P(x^\prime, t+\tau|x, t)$ is the transition probability to some new
state $x^\prime$ at time $t+\tau$ from the initial state $x$ at time
$t$, then the following integral equation determines all subsequent
evolution of the distribution $f(x)$:
\begin{displaymath}
  f(x, t+\tau) = \int dx^\prime  P(x, t+\tau|x^\prime, t) f(x^\prime, t).
\end{displaymath}
By expanding the transition probability in its moments of $x-x^\prime$
for small $\tau$, one may derive a differential equation of the form:
\begin{equation}
  {\partial f(x, t)\over\partial t} = 
  \sum_{n=1}^\infty \left(-{\partial\over\partial x}\right)^n
      D^{(n)}(x, t) f(x, t)
\label{eq:km}
\end{equation}
where the coefficients $D^{(n)}$ are proportional to the
time-derivative of the moments in transition probability.  This is
known as the {\em Kramers-Moyal expansion}.

For galaxy evolution, the state variable $x$ in the equations above is
replaced with the phase-space six vector.  Since we are interested in
long-term evolution, we may orbit average the evolution equation.
Writing the phase space as actions and angles turns the orbit average
into an angle average.  For any given noise process, we can solve for
the change in actions of any orbit in the galaxy using a perturbative
approach and similarly derive the change in the phase-space
distribution function as described in Weinberg
(1998).  Evaluation of these quantities lead
directly to the moments needed for the Fokker-Planck-type evolution
equation.  The most subtle aspect of the development below is
enforcing a consistent time ordering.  Implicit in the angle averaging
is a short dynamical time scale and a long evolutionary time scale.
The stochastic perturbations are on short time scales and
instantaneous from the evolutionary point of view.  Although this
approximation may seem restrictive and only marginally true for some
scenarios of interest, it yields results in good agreement with n-body
simulation in the case of globular cluster evolution and the closely
related case of self-gravitating fluctuations explored in Weinberg
(1998).

After deriving the evolution equation below, we work out the details of
the Fokker-Planck coefficients for two examples in \S\ref{sec:noise}:
transient distortions such as fly-by encounters and quasi-periodic
distortions such as orbiting super-massive black holes.  The
consequences of both of these will be explored in Paper 2.

\subsection{Derivation of the Fokker-Planck equation}
\label{sec:fp}

The main advantage of the Kramers-Moyal expansion is that the noise
process appears explicitly in an initial-value form.  Because the
response of the galaxy to a stochastic event can be straightforwardly
computed in perturbation theory or by n-body simulation, we can
compute the noise-driven evolution for a wide variety of scenarios.
For this reason, it is better suited to treating general stochastic
noise than the Master equation, even though the two approaches are
formally equivalent.

Although the Kramers-Moyal expansion has an infinite number of terms in
general (cf. eq. \ref{eq:km}), the Pawula Theorem (Pawula
1967) shows that consistency demands that the
expansion either stops after two terms and takes the standard
Fokker-Planck form or must have an infinite number of terms.  For the
stellar dynamical case considered here, the series truncates after two
terms.  The approach approach sketched here is clearly described by
Risken (1989, Chap.  4).

As outlined in \S\ref{sec:overview}, one begins the derivation with
the definition of transition probability:
\begin{equation}
  f(\bI, t+\tau) = \int d\bI^\prime\,
  P(\bI, t+\tau | \bI^\prime, t) f(\bI^\prime, t)
  \label{eq:cprob}
\end{equation}
where an average over the rapidly oscillating angles is implied and
$P$ is the conditional probability that a state has $\bI$ at time
$t+\tau$ if it has $\bI^\prime$ at time $t$ initially.  A Taylor
series expansion of the integrand in $\Delta\equiv\bI^\prime -
\bI$ followed by a change of variables and integration over
$\Delta$.  In the limit $\tau\rightarrow0$ this expansion leads
directly to
\begin{equation}
  {\partial f(\bI, t+\tau)\over\partial t} = 
  \sum_{n=1}^\infty \left(-{\partial\over\partial \bI}\right)^n
  D^{(n)}(\bI, t)   f(\bI, t) .
\end{equation}
Note that $P$ is the probability of a change in $\bI$ due to
stochastic events.  Therefore, the formal time derivatives in the
expansion and phase space integral is better considered as the limit
for small $\tau$ (but for $\tau$ greater than a dynamical time) of the
ensemble average of stochastic events.  If we let $\bxi$ be the
stochastic value of $\bI$, the expansion coefficients describing the
stochastic variables $\bxi$ is:
\begin{equation}
D^{(n)}(x, t) = \left.{1\over n!}\lim_{\tau\rightarrow0}{1\over\tau}
\langle[\bxi(t+\tau) - \bI]^n\rangle\right|_{\bxi(t)=\bI}.
\label{eq:kramers}
\end{equation}
In $N$ dimensions, this becomes:
\begin{equation}
D^{(n)}_{j_1,j_2,\ldots,j_n}(\bI, t) = 
{1\over n!}\lim_{\tau\rightarrow0}{1\over\tau} 
M^{(n)}_{j_1,j_2,\ldots,j_n}(\bI, t, \tau)
\label{eq:diff}
\end{equation}
where the moments $M^{(n)}$ are
\begin{equation}
M^{(n)}_{j_1,j_2,\ldots,j_n}(\bx, t, \tau)
=
\left\langle
(\barI_{j_1}-I_{j_1})(\barI_{j_2}-I_{j_2})\cdots(\barI_{j_n}-I_{j_n})
\right\rangle.
\label{eq:moms}
\end{equation}
The angle brackets denote the integration over the conditional
probability of obtaining the state variable $\barI_j$ at time $t+\tau$
given $I_j$ at time $t$.

We will see below that the response function guarantees the that the
Kramers-Moyal expansion terminates after two terms in the limit of
weak perturbations and will assume this here.  This is consistent with
our intuition that repetitive weak, stochastic excitation in a galaxy
is a Markov process.  The evolution equation is then the Fokker-Planck
equation in standard form:
\begin{equation}
{\partial f(\bI, t)
  \over
  \partial t} =
{\bf L}_{FP}(\bI, t) f(\bI, t)
\label{eq:FP}
\end{equation}
where
\begin{equation}
{\bf L}_{FP}(\bI, t) =
-{\partial\over\partial I_i} D^{(1)}_i(\bI, t) + 
{\partial^2\over\partial I_i\partial I_j} D^{(2)}_{ij}(\bI, t).
\end{equation}

We see that the natural coordinates for the Boltzmann equation are
action-angle variables.  For a collisionless equilibrium, the actions
are constant and the angles advance at constant rate.  In deriving
this Fokker-Planck equation, one averages over orbital periods in
favor of following the evolution over longer time scales.  The
distribution function is then a function of actions alone, $f=f(\bI)$.

However, we will concentrate on the evolution of haloes modeled as a
collisionless spherical distribution for application in Paper 2 and
adopt the traditional $E, J, J_z$ or $E, \kappa\equiv J/J_{max}(E),
\cos\beta\equiv J_z/J$ phase-space variables for computational
purposes.  In addition, we will not consider any processes with a
preferred axis so we can average equation (\ref{eq:FP}) over $\beta$
with no loss of information.  If we are willing to restrict ourselves
to an isotropic distribution, we can average over $\kappa$ to yield a
$1+1$ dimensional Fokker-Planck equation in time and energy, $E$; this
is described in \S\ref{sec:fpavg} below.  Because the diffusion
coefficients depend on the distribution function, the Fokker-Planck
equation in (\ref{eq:FP}) is non-linear, just as for globular cluster
evolution.  However, it is straightforward to solve this numerically
by iteration.

To derive the coefficients $D^{(1)}$ and $D^{(2)}$, we will use the
method described in Weinberg (1998, hereafter Paper
1).  To summarize, we represent distortions in
the structure of halo in a biorthogonal basis.  Any distortion can be
summarized then by a set of coefficients in three indices.  Because
large spatial scales are most important in understanding global
evolution, we can truncate this expansion and still recover most of
the power.  Moreover, we can analytically compute the self-gravitating
response of the halo to some arbitrary perturbation as previously
described.  This development gives us $\xi(t)$ for all phase-space
variables (cf.  eq.  \ref{eq:kramers}) and the appropriate ensemble
averages give us the required diffusion coefficients.  For example, if
one uses the same biorthogonal basis in an n-body simulation of a
desired transient process, the time series of coefficients can be used
directly to derive the coefficients $D^{(1)}$ and $D^{(2)}$ after
removing the time-invariant (DC) component which corresponds to the
equilibrium background.  We will consider point mass perturbers in
\S\ref{sec:pointmass} and transient perturbers (dwarf galaxies,
decaying substructure, spreading debris trails) in
\S\ref{sec:shrapnel}.

\subsection{Averaged Fokker-Planck equation}
\label{sec:fpavg}

A number of authors have described transformation of the multivariate
Fokker-Planck equation (Rosenbluth et al. 1957, Risken 1989).  The
approach is the familiar one: write the equation in terms of scalars,
covariant and contravariant vectors and tensors and covariant
derivatives only.  In the first case, the authors use the Jacobian of
the coordinate transformation as a metric and in the second, the
authors use the diffusion matrix.  We will use the first case here.
Denote the Jacobian of the coordinate transformation as $J$.  Under a
change of coordinates, one can show after a fair bit of algebra that
the advection and diffusion terms transform as
\begin{eqnarray}
  D^\prime_k &=& {\partial I^\prime_k\over\partial I_i} D_i
   + {\partial^2 I^\prime_k\over\partial I_i\partial I_j}D_{ij}, \\
  D^\prime_{kl} &=& {\partial I^\prime_k\over\partial I_i}
  {\partial I^\prime_l\over\partial I_j} D_{ij}.
\label{eq:diftrans}
\end{eqnarray}
The phase-space distribution function transforms as $f^\prime(\bI)
= J f(\bI)$ (cf. Risken 1989) and in the new variables, the
equation takes the standard Fokker-Planck form:
\begin{equation}
  {\partial f^\prime(\bI, t)\over\partial t} =
  \left[-{\partial\over\partial I_k^\prime}D_k^\prime +
    {\partial^2\over\partial I_k^\prime \partial I_l^\prime}
  D_{kl}^\prime\right] f^\prime(\bI, t).
\label{eq:fptrans}
\end{equation}

Now let $\bI^\prime = (E, \kappa, \cos\beta)$.  Assuming that the
distribution function $f$ is time-independent and non-zero, we may
integrate equation (\ref{eq:fptrans}) over $\kappa$ and $\cos\beta$.
Since both of these variables have a bounded domain, the flux through
their boundaries must vanish, leaving a single flux term:
\begin{equation}
{\partial\langle f^\prime\rangle\over\partial\,t} = 
{\partial\over\partial E}\left\langle
-D_E f^\prime + {\partial \over\partial I^j}\left(D_{Ej}
  f^\prime\right)\right\rangle_{iso} 
\end{equation}
where the angle brackets denote integration over $\kappa$ and
$\cos\beta$ and sum over $j$ denote the sum over all three variables.
The isotropically averaged Fokker-Planck equation is then
\begin{equation}
{\partial {\bar f}(E)\over\partial\,t} = 
{\partial\over\partial E}\left[
-\langle D_E\rangle_{iso} {\bar f}(E) + {\partial \over\partial
  E}\left(\langle D_{EE}\rangle_{iso} {\bar f}(E)\right)\right]
\label{eq:fpeavg}
\end{equation}
where $\langle D_E\rangle_{iso}$ and $\langle D_{EE}\rangle_{iso}$ are the
isotropically averaged diffusion coefficients:
\begin{eqnarray}
\left\langle{D_E\atop D_{EE}}\right\rangle &=& {J_{max}^2(E)\over f(E)}
\int d\kappa\,d(\cos\beta)\,\kappa \left\{{D_E\atop D_{EE}}\right\}
f(E,\kappa,\beta) \\ \noalign{\leftline{where}}
{\bar f}(E) &=& J_{max}^2(E)
\int d\kappa\,d(\cos\beta)\,\kappa  f(E,\kappa,\beta) \\
\noalign{\leftline{and the phase-space volume is}}
P(E) &\equiv& J_{max}^2(E)
\int d\kappa\,d(\cos\beta)\,\kappa.
\end{eqnarray}
Note that the standard notation in the globular cluster literature is
$f(E)={\bar f}(E)/P(E)$.

\section{Noise models}
\label{sec:noise}

As described in \S\ref{sec:derive}, we represent distortions in the
structure of halo in a biorthogonal basis.  Any distortion can then be
summarized by a set of coefficients.  Because large spatial scales are
most important in understanding global evolution, we can truncate this
expansion and still recover most of the power.  Internal and therefore
quasi-periodic distortions contribute at a discrete spectrum of
frequencies.  Paper 1, \S2 (see eqns. 21-22 in Paper 1 for the final
result) derives the response of a halo to a point perturbation at a
single frequency.  Similar arguments lead to an expression for a
continuous spectrum of perturbation frequencies.  In the latter case,
one computes the response of the stellar system to each frequency in
the spectrum and then sums over all frequencies.  We will begin with
the development common to both cases.

The goal is calculation of the coefficients defined by equations
(\ref{eq:kramers}) and (\ref{eq:diff}).  We begin by determining these
coefficients for action variables and transform to $(E, \kappa,
\cos\beta)$ in the end.  Because orbits in the equilibrium phase space
are quasi-periodic and representable as fixed actions and constantly
advancing angles, any perturbed quantity can be represented as a
Fourier series in angles with coefficients depending on actions.
Following Paper 1, the perturbed Hamiltonian is
\begin{eqnarray}
  H(\bI, {\bf w}) &=& H_o(\bI) + H_1(\bI, {\bf w}) \\
&=& H_o(\bI) + \sum_{\bf l} H_{1{\bf l}}(\bI) e^{{\bf
    l}\cdot{\bf w}} \\ \noalign{\leftline{where}}
H_{1{\bf l}}(\bI) &=& 
\sum_{l=0}^\infty\sum_{m=-l}^l
\sum_j Y_{ll_2}(\pi/2,0)r^l_{l_2m}(\beta)W^{l_1\,j}_{ll_2m}(\bI)
a^{lm}_j(t) \\
\label{eq:h1}
&=& 
\sum_{l=0}^\infty\sum_{m=-l}^l
\sum_j Y_{ll_2}(\pi/2,0)r^l_{l_2m}(\beta)W^{l_1\,j}_{ll_2m}(\bI)
\int^\infty_{-\infty}d\omega e^{i\omega t} \sum_k \left[{\cal
  M}^{lm}_{jk}(\omega) + \delta_{jk}\right] b^{lm}_k(\omega) \nonumber \\
\label{eq:h1a}
\end{eqnarray}
where ${\bf l}={l_1, l_2, l_3}$ is a triple of integers,
$r^l_{ij}(\beta)$ and $W^{l_1\,j}_{ll_2l_3}(\bI)$ are the rotation
matrices and gravitational potential transforms defined in Paper 1.
The time dependence of the coefficients describing the response,
$a^{lm}_j(t)$, is represented as its Fourier transform.  This allows
each frequency to be treated separately.  The response matrix ${\cal
  M}$ describes the reaction of the galaxy to the perturbation; so the
entire response is the sum of both the response and direct forcing,
${\cal M}^{lm}_{jk} + \delta_{jk}$.  We may integrate the
equations of motion directly to evaluate $\bI(\tau+t)$.
Hamilton's equations yield
\begin{equation}
  {\dot I_j} = -{\partial H\over\partial w_j} = -i \sum_{\bf l}
    l_j H_{1{\bf l}}(\bI) e^{{\bf l}\cdot{\bf w}}
    \label{eq:idot}
\end{equation}
and therefore we have
\begin{equation}
  \Delta I_j(t+\tau) \equiv  I_j(t+\tau) - I_j(t) = 
  \int^{t+\tau}_t dt {\dot I_j}(t).
\end{equation}

The evolution of perturbed distribution function in time follows from 
the linearized collisionless Boltzmann equation and the total time
derivative for a Hamiltonian system:
\begin{equation}
  {\dot f}_1 \equiv {\partial f_1\over\partial t} + [f_1, H] = {\partial f_1\over\partial t} + {\partial
    H_0\over\partial\bI}\cdot{\partial f_1\over\partial{\bf w}} = 
    {\partial H_1\over\partial{\bf w}}\cdot{\partial f_0\over\partial\bI}.
\end{equation}
Analogous to the development above for $H_1(t)$, we have
\begin{equation}
  {\dot f}_1(\bI, {\bf w}, t) = \sum_{\bf l} e^{i{\bf
      l}\cdot{\bf w}} i{\bf l}\cdot{\partial f_0\over\partial\bI}
  H_{1{\bf l}}(\bI, t)
\label{eq:f1}
\end{equation}
and therefore
\begin{equation}
  f_1(\bI, {\bf w}, t+\tau) = 
  \sum_{\bf l} e^{i{\bf l}\cdot{\bf w}} 
  i{\bf l}\cdot{\partial f_0\over\partial\bI}
  \int^{t+\tau}_t dt^\prime H_{1{\bf l}}(\bI, t^\prime)
\label{eq:f1a}
\end{equation}

To evaluate equation (\ref{eq:diff}) we need the first and second
order action moments defined by equation (\ref{eq:moms}). We assume,
by adopting the time-asymptotic response matrix ${\cal M}^{lm}$ in
deriving that $f_1$ and $H_1$ above, that $\tau$ is larger than
intrinsic dynamical times, consistent with the ordering of our slow
and fast time scales.  Previous work, including comparison to n-body
simulations, suggests that this is a very good approximation for time
scales longer than several crossing times.  The response to the
distortion induces a shift in the actions $\bI$ and the overall
response causes a change in the distribution function.  This is
represented in the matrix equation defined by equation (\ref{eq:h1})
for an external perturbation described by the coefficients
$b^{lm}_j(t)$.  The fiducial stochastic variables are the coefficients
$b^{lm}_j(t)$ themselves.

The overall conditional probability required in equation
(\ref{eq:cprob}) and the following development for the moments
therefore has two contributions.  First, the response of the galaxy
changes the underlying distribution and consequently the probability
of obtaining a given final state.  Second, the resonant coupling
changes the action of an orbit at a particular point in phase space.
Altogether we have
\begin{equation}
  P(\bI^\prime, t+\tau | \bI, t) = \left(1 + {f_1({\bf
  I}, t)\over f_0(\bI)}\right) 
{\bf\delta}\left(\bI^\prime - \bI - \int^{t+\tau}_t dt^\prime
  \Delta{\bf{\dot I}}(t^\prime)\right).
\end{equation}

The first- and second-order moments are proportional to the square of
the perturbation coefficients $b^2$ and are therefore second-order in
the perturbation amplitude.  With additional work, one can show that
the next contributing order is proportional to $b^4$ and therefore
relatively negligible.  We will denote the ensemble average of the
fluctuating coefficients which we will denote as angle brackets, e.g.:
$\langle b^{lm}_j(t_1) b^{lm}_j(t_2) \cdots b^{lm}_j(t_n) \rangle$.
Note that $\left\langle b^{lm}_k (t_1) b^{\ast lm}_k (t_2)
\right\rangle$ is related to the density correlation function:
\begin{eqnarray}
\left\langle b^{lm}_k (t_1) b^{\ast lm}_k (t_2) \right\rangle
&=& \int d^3r_1 \int d^3r_2\,
Y^\ast_{lm}(\theta_1,\phi_1)
Y_{lm}(\theta_2,\phi_2) u^{\ast\,lm}_j(r_1) u^{lm}_k(r_2) 
\left\langle \rho({\bf r_1}, t_1) \rho({\bf r_2}, t_2) \right\rangle.
\end{eqnarray}
Assuming that the process causing fluctuations is independent of time
(e.g. a {\em stationary} process) we can write $\left\langle \rho({\bf
    r_1}, t_1) \rho({\bf r_2}, t_2) \right\rangle = C({\bf r_1}, {\bf
  r_2}, t_1-t_2)$.  The quantity $\left\langle b^{lm}_k (t_1) b^{\ast
    lm}_k (t_2) \right\rangle = \left\langle b^{lm}_k (0) b^{\ast
    lm}_k (t_1-t_2) \right\rangle$ describes the correlation of random
variables $b^{lm}_j$ as a function of time.  The limit $t_1\rightarrow
t_2$ gives the mean-squared fluctuation amplitude and was explored and
compared to n-body simulations in Paper 1.

We can now use equations (\ref{eq:h1a}), (\ref{eq:idot}) and
(\ref{eq:f1a}) to evaluate the moments in equation (\ref{eq:moms}).
After explicit substitution and averaging over angles, we have
\begin{eqnarray}
  \left\langle \Delta I_j(t+\tau) \right\rangle
  &=& 
  \left({1\over2\pi}\right)^2
  \int^\infty_{-\infty}d\omega_1
  \int^\infty_{-\infty}d\omega_2
  \left\langle
    i{\bf l}\cdot{\partial\ln f_o\over\partial\bI}
    \int^{t+\tau}_t dt_1 e^{i\omega_1 t}
    \int^{t+\tau}_t dt_2 e^{i\omega_2 t}
  \right.
  \times \nonumber \\
  &&
  \sum_\mu Y_{ll_2}(\pi/2,0)r^l_{l_2m}(\beta)W^{l_1\,\mu}_{ll_2m}({\bf
    I}) \sum_r {\cal M}^{lm}_{\mu r}(\omega_1)\, b^{lm}_r(t_1) 
  \times \nonumber
  \\
  &&
  \left.
    i l_j \sum_\nu
    Y_{ll_2}(\pi/2,0)r^l_{l_2m}(\beta)W^{\ast l_1\,\nu}_{ll_2m}(\bI)
    \sum_s {\cal M}^{\ast lm}_{\nu s}(\omega_2)\, b^{\ast lm}_s(t_2)
  \right\rangle \nonumber \\
  &=& 
  -l_j{\bf l}\cdot{\partial\ln f_o\over\partial\bI}
  \left|Y_{ll_2}(\pi/2,0)\right|^2 
  \left({1\over2\pi}\right)^2
  \int^{\infty}_{-\infty}d\omega_1
  \int^{\infty}_{-\infty}d\omega_2
  \times \nonumber \\
  &&
  \sum_{\mu\nu}\sum_{rs}
  r^l_{l_2m}(\beta)r^{\ast l}_{l_2m}(\beta)
  W^{l_1\,\mu}_{ll_2m}(\bI) 
  W^{\ast l_1\,\nu}_{ll_2m}(\bI) 
  {\cal M}^{lm}_{\mu r}(\omega_1)
  {\cal M}^{\ast lm}_{\nu s}(\omega_2)
  \times \nonumber
  \\
  &&
  \int^{t+\tau}_t dt_1
  \int^{t+\tau}_t dt_2
  e^{i(\omega_1t_1+\omega_2t_2)}
  \left\langle
    b^{lm}_r (\omega_1)
    b^{\ast lm}_s (\omega_2)
  \right\rangle .
  \label{eq:mom1exp}
\end{eqnarray}
The expression for $\left\langle \Delta I_j(t+\tau) \Delta I_k(t+\tau)
\right\rangle$ is nearly the same, with ${\bf l}\cdot d\ln f_o/d\bI$ in
equation (\ref{eq:mom1exp}) replaced with $-l_k$.  Although cumbersome
in appearance, all quantities in this expression are straightforwardly
computed.  In particular, the rotation matrices, $
r^l_{l_2m}(\beta)$, have closed-form analytic expressions and the
response matrices, ${\cal M}^{lm}_{\mu r}(\omega)$, have elements that
can be evaluated by quadrature.  For a specific stochastic process,
all that remains is to evaluate the Fourier transform of the density
correlation function.  We will do this below for quasi-periodic and
transient noise sources.

\subsection{Orbiting point mass perturbers}
\label{sec:pointmass}

For a halo of black holes, we may assume that the perturbers are point
masses.  In other words, the density $\rho$ is a sum of delta
functions.  Expanding the distribution for a single black home in an
action-angle series gives
\begin{eqnarray}
  b^{lm}_j(t) &=& \sum_{\bf l} Y_{ll_2}(\pi/2,0)r^l_{l_2m}(\beta)
  W^{\ast l_1\,\nu}_{ll_2m}(\bI) e^{i{\bf l}\cdot{{\bf w}(t)}}
    \label{eq:btbt}
\\ \noalign{\leftline{where ${\bf w}(t) = {\bf w}_o +
    \bOmega t$ and after Fourier transforming in time, we find}}
  b^{lm}_j(\omega) &=& \sum_{\bf l} Y_{ll_2}(\pi/2,0)r^l_{l_2m}(\beta)
  W^{\ast l_1\,\nu}_{ll_2m}(\bI) e^{i{\bf l}\cdot{\bf w}_o}
  2\pi\delta\left(\omega - {\bf l}\cdot\bOmega\right)
  \label{eq:bwbw}
\end{eqnarray}
As in Paper 1, we assume that orbits of individual particles are
uncorrelated.  The particle wakes do in fact give rise to correlations
but this is of higher order in $1/N$ in the BBGKY expansion than the
lowest-order effect we will consider here (cf. Gilbert
1969).  The number density of particles at $\bI_1,
{\bf w}_1$ at time $t_1$ and at ${\bf I}_2, {\bf w}_2$ at time $t_2$
is
\begin{equation}
        {\cal P}(\bI_1, {\bf w}_1, t_1; \bI_2, {\bf
        w}_2, t_2) = {\cal P}(\bI_1, {\bf w}_1)
        \delta(\bI_1 - \bI_2)
        \delta({\bf w}_1 - {\bf w}_2 + {\bOmega}(\bI_1)(t_2-t_1))
        \label{eq:dist2}
\end{equation}
where ${\cal P}(\bI, {\bf w})$ is the equilibrium particle
distribution with
\begin{equation}
N=\int d^3I d^3w {\cal P}(\bI, {\bf w}).
\end{equation}
Direct substitution demonstrates that equation (\ref{eq:dist2}) solves
the Liouville equation with the initial condition $\bI_2 =
\bI_1$ and ${\bf w}_2 = {\bf w}_1$ at $t=t_1$.  Similarly,
integrating equation (\ref{eq:dist2}) over all coordinates gives $N$.
The ensemble average $\left\langle b^{lm}_r (t_1) b^{\ast lm}_s (t_2)
\right\rangle$ is then the average of $b^{lm}_r (t_1) b^{\ast lm}_s
(t_2)$ the distribution given by equation (\ref{eq:dist2}).

We now apply this two-particle distribution function to explicitly
evaluate the ensemble average in equation (\ref{eq:mom1exp}).  
The ensemble average here implies an average of possible distributions
of point masses consistent with some underlying distribution.
Therefore
\begin{eqnarray}
  \left\langle
    b^{lm}_r (\omega_1)
    b^{\ast l^\prime m^\prime}_s (\omega_2)
  \right\rangle &=& \left\langle
    \sum_{\bf l} \delta_{mm^\prime}
    Y_{ll_2}(\pi/2,0) Y^\ast_{l^\prime l_2}(\pi/2,0) 
    r^l_{l_2m}(\beta) r^{\ast l^\prime}_{l_2}(\beta)
    \times \right. \nonumber \\
  && \left.
    W^{l_1\,\nu}_{ll_2m}(\bI) 
    W^{\ast l_1\,\mu}_{l^\prime l_2m}(\bI) 
    (2\pi)^2 \delta(\omega_1+\omega_2)
  \delta(\omega_1- {\bf l}\cdot\bOmega) \rule{0pt}{20pt} \right\rangle .
\nonumber \\
  \label{eq:b2pmass}
\end{eqnarray}
and
\begin{equation}
  \left\langle
    b^{lm}_r (t_1)
    b^{\ast l^\prime m}_s (t_2)
  \right\rangle =
  \left({1\over2\pi}\right)^2
  \int^\infty_{-\infty} d\omega_1
  \int^\infty_{-\infty} d\omega_2 e^{i(\omega_1 t_1 - \omega_2 t_2)}
  \left\langle
    b^{l^\prime m}_r (\omega_1)
    b^{\ast lm}_s (\omega_2)
  \right\rangle.
  \label{eq:t2pmass}
\end{equation}
Integration over angles identifies ${\bf l}$ with ${\bf l}^\prime$ and
therefore $m=l_3=l_3^\prime=m^\prime$.  Now using equations
(\ref{eq:btbt}), (\ref{eq:bwbw}), (\ref{eq:dist2}) to evaluate the
average we find that the final part of expression (\ref{eq:mom1exp})
becomes
\begin{eqnarray}
  &&
  \left({1\over2\pi}\right)^2
  \left\langle
    \int^\infty_{-\infty}d\omega_1
    \int^\infty_{-\infty}d\omega_2
    e^{i(\omega_1t_1+\omega_2t_2)}
    b^{lm}_r (\omega_1)
    b^{\ast lm}_s (\omega_2)
  \right\rangle = \nonumber \\ &&
  (2\pi)^3 \sum_{\bf l} \int d^3I f_o(\bI)
  e^{i{\bf l}\cdot\bOmega(\bI)(t_1-t_2)}
  |Y_{ll_2}(\pi/2,0)|^2
  |r^l_{l_2m}(\beta)|^2
  W^{l_1\,r}_{ll_2m}(\bI) W^{\ast l_1\,s}_{ll_2m}(\bI)
  \nonumber \\
\end{eqnarray}
where the integration over $\omega_1$ and $\omega_2$ has become a sum
over discrete frequencies denoted ${\bf l}$ and we have exploited the
orthogonality of rotation matrices:
\begin{equation}
\int d\beta \sin(\beta) r^l_{\mu\nu}(\beta)\, r^{l^\prime}_{\mu\nu}(\beta) = 
{2\over 2l+1}\delta_{l\,l^\prime}
\end{equation}
(Edmonds 1960).

We now substitute this development into equation (\ref{eq:mom1exp})
and find
\begin{eqnarray}
  \left\langle \Delta I_j(t+\tau) \right\rangle
  &=& 
  -{1\over f_0(\bI)}
  l_j{\bf l}\cdot{\partial f_o\over\partial\bI}
  \left|Y_{ll_2}(\pi/2,0)\right|^2 
  \sum_{\mu\nu}\sum_{rs}
  r^l_{l_2m}(\beta)r^{\ast l}_{l_2m}(\beta)
  W^{l_1\,\mu}_{ll_2m}(\bI) 
  W^{\ast l_1\,\nu}_{ll_2m}(\bI) 
  \times \nonumber \\ && 
\left\{
  (2\pi)^3 \sum_{\bf l} \int d^3I f_o(\bI)
  |Y_{ll_2}(\pi/2,0)|^2
  r^l_{l_2m}(\beta) r^l_{l_2m}(\beta)
  W^{l_1\,r}_{ll_2m}(\bI) W^{\ast l_1\,s}_{ll_2m}(\bI)
  {\cal M}^{lm}_{\mu r}({\bf l}\cdot\bOmega(\bI))
  {\cal M}^{\ast lm}_{\nu s}({\bf l}\cdot\bOmega(\bI))
  \right.\times \nonumber \\
  && \left.
  \int^{t+\tau}_t dt_1
  \int^{t+\tau}_t dt_2
  e^{i{\bf l}\cdot\bOmega(\bI)(t_1-t_2)} \right\}
  .
  \label{eq:mom1exp2}
\end{eqnarray}
As described generally above, the term in $\{\cdot\}$ is the temporal
correlation function for the response to the point masses fluctuations
and only depends on the time difference $t_1-t_2$.  The correlation is
finite and lasts some order-unity number of dynamical times.  Our
diffusion calculation is in the regime $\tau\gg1/\Omega$.  We may
change the double time integration from variables $t_1, t_2$ to
$T=(t_1+t_2)/2, \tau=t_1-t_2$.  The integral over $\tau$ gives a delta
function $\delta({\bf l}\cdot\bOmega(\bI))$ and the integral over $T$
gives $\tau$.  The delta function implies that only orbits with
commensurate frequencies will give rise to secular changes.
Physically, the disturbance must present an asymmetric force
distribution to cause a secular change in the actions of dark matter
orbits.  If the orbital frequencies are not commensurate, the
long-term average of the perturbing force will be axisymmetric.  For
orbiting black holes, the ratio $\Omega_1/\Omega_2$ ranges from 1 to 2
as the halo potential varies from that for point mass to homogeneous
core.  Therefore for most systems, no commensurabilities are available
for harmonic orders $l=1,2$.  In other words, a resonant $l=1$ orbit
will look like a Keplerian orbit and a resonant $l=2$ orbit will look
like a bisymmetric oval (e.g.  stationary bar orbit) and neither exist
in any significant measure in most extended stellar systems.

Returning to equation (\ref{eq:kramers}), we can now read off our
diffusion coefficients in action variables:
\begin{eqnarray}
  D^{(1)}_j(\bI, t) &=& \lim_{\tau\rightarrow0}
  {\left\langle \Delta I_j(t+\tau) \right\rangle \over \tau} \nonumber \\
  &=&
  -{1\over f_0(\bI)}
  l_j{\bf l}\cdot{\partial f_o\over\partial\bI}
  \left|Y_{ll_2}(\pi/2,0)\right|^2 
  \sum_{\mu\nu}\sum_{rs}
  r^l_{l_2m}(\beta)r^{\ast l}_{l_2m}(\beta)
  W^{l_1\,\mu}_{ll_2m}(\bI) 
  W^{\ast l_1\,\nu}_{ll_2m}(\bI) 
  \times \nonumber
  \\
  && \left\{
  (2\pi)^3 \sum_{\bf l} \int d^3I f_o(\bI)
  |Y_{ll_2}(\pi/2,0)|^2 
  r^l_{l_2m}(\beta) r^l_{l_2m}(\beta) 
  \times \right. \nonumber \\ && \left.
  W^{l_1\,r}_{ll_2m}(\bI) W^{\ast l_1\,s}_{ll_2m}(\bI)
  {\cal M}^{lm}_{\mu r}({\bf l}\cdot\bOmega(\bI))
  {\cal M}^{\ast lm}_{\nu s}({\bf l}\cdot\bOmega(\bI))
  2\pi\delta\left({\bf l}\cdot\bOmega(\bI)\right)
  \right\}, \nonumber \\
  \\
  D^{(2)}_{jk}(\bI, t) &=& \lim_{\tau\rightarrow0}
  {\left\langle \Delta I_j(t+\tau) \Delta I_k(t+\tau) \right\rangle \over 2\tau} \nonumber \\
  &=&
  {l_jl_k\over 2}
  \left|Y_{ll_2}(\pi/2,0)\right|^2 
  \sum_{\mu\nu}\sum_{rs}
  r^l_{l_2m}(\beta)r^{\ast l}_{l_2m}(\beta)
  W^{l_1\,\mu}_{ll_2m}(\bI) 
  W^{\ast l_1\,\nu}_{ll_2m}(\bI) 
  \times \nonumber
  \\
  && \left\{
  (2\pi)^3 \sum_{\bf l} \int d^3I f_o(\bI)
  |Y_{ll_2}(\pi/2,0)|^2 
  r^l_{l_2m}(\beta) r^l_{l_2m}(\beta) 
  \times \right. \nonumber \\ && \left.
  W^{l_1\,r}_{ll_2m}(\bI) W^{\ast l_1\,s}_{ll_2m}(\bI)
  {\cal M}^{lm}_{\mu r}({\bf l}\cdot\bOmega(\bI))
  {\cal M}^{\ast lm}_{\nu s}({\bf l}\cdot\bOmega(\bI))
  2\pi\delta\left({\bf l}\cdot\bOmega(\bI)\right)
  \right\}. \nonumber \\
\label{eq:mom2exp}
\end{eqnarray}
Note that the limit $\tau\rightarrow0$ is taken in the sense that
$\tau$ is small compared to the evolutionary time scale due to the
fluctuations but large compared to the dynamical time.  The time
dependence in the diffusion coefficients reminds us that the
underlying equilibrium distribution $f_o(\bI)$ changes on an
evolutionary time scale but, for the purposes of computation, is held
fixed on a dynamical time scale.  The integrals may be simplified by
noting that $d^3I=dE dJ J d(\cos\beta)/\Omega_1(E,J)$, we can do the
integral in $\beta$ using the orthogonality of the rotation matrices
as previously described.  For a given {\em background} distribution
function $f_o(\bI)$, the term in curly brackets can be computed once
and for all since they are independent of the local value of the
actions.

\subsection{Transient processes}
\label{sec:shrapnel}

In the previous application for orbiting point masses, we saw that
only harmonics with $l\ge3$ were effective at driving evolution.
Here, we describe the results of bombarding the galaxy isotropically
with bits of mass. This is an idealization of interactions during the
epoch of galaxy evolution.  The perturbations are {\it shots}, not
quasi-periodic, and therefore lead to a continuous spectrum of
perturbation frequencies.  This case and the previously considered
point-mass case represent two extremes.  For example, a decaying orbit
will have a set of broad peaks and a low-frequency continuum.

The expansion coefficients for the perturbation are straightforward
for the mass inside of the galaxy:
\begin{equation}
  b^{lm}_i(t) = \int d^3r\,Y_{lm}^\ast(\theta, \phi) u^{\ast\,lm}_i(r) 
  \delta\left({\bf r} - {\bf r}(t)\right) = 
  Y_{lm}^\ast(\theta(t), \phi(t)) u^{\ast\,lm}_i(r(t)).
  \label{eq:porbit1}
\end{equation}
If the mass is outside the galaxy, we use the multipole expansion
with the density rather than the potential member of the biorthogonal
pair to evaluate the coefficients:
\begin{eqnarray}
  b^{lm}_i(t) &=& 
  -{1\over4\pi}
  \int d^3r\,Y_{lm}^\ast(\theta, \phi) d^{\ast\,lm}_i(r) {1\over |{\bf
  r} - {\bf r}(t)|} \nonumber \\
&=&
  -{1\over4\pi}
  \int d^3r\,Y_{lm}^\ast(\theta, \phi) d^{\ast\,lm}_i(r)
  \sum_l\sum_{m=-l}^l {4\pi\over2l+1} {r_<^l\over r_>^{l+1}}
  Y_{lm}^\ast(\theta(t), \phi(t)) Y_{lm}^\ast(\theta, \phi) \nonumber \\
&=&
  - {1\over2l+1} Y_{lm}^\ast(\theta(t), \phi(t))
  \int dr r^2\, d^{\ast\,lm}_i(r) {r^l\over r(t)^{l+1}}
  \label{eq:porbit2}
\end{eqnarray}

The Fourier transform needed for equation (\ref{eq:mom1exp}) is most
easily done assuming the perturber is in the equatorial plane.  We
denote the Fourier transform of $b^{lm}_j(t)$ as ${\hat
  b}^{lm}_j(\omega)$.  In practice, we will perform the transform
numerically by FFT.  Note that equations (\ref{eq:porbit1}) and
(\ref{eq:porbit2}) describe the time dependence in coefficients for
any trajectory $r(t), \theta(t), \phi(t)$. A cloud of points and more
generally any phase-space distortion yielding $b^{lm}_i(t)$ or
coefficients at discrete times from an n-body simulation are possible
input to the FFT.  Now, we can evaluate the final line in equation
(\ref{eq:mom1exp}) by changing coordinates from $t_1, t_2$ to
$T=(t_1+t_2)/2, \tau=t_1 - t_2$.  We have
\begin{displaymath}
  e^{i(\omega_1 t_1 + \omega_2 t_2)} =
  e^{i(\omega_1 [T+\tau/2] + \omega_2[T-\tau/2])} =
  e^{i(\omega_1 + \omega_2)T} e^{i(\omega_1 - \omega_2)\tau/2}.
\end{displaymath}
Using this in the last line of equation (\ref{eq:mom1exp}) we find
\begin{eqnarray}
  \int^{t+\tau}_t dt_1
  \int^{t+\tau}_t dt_2
  e^{i(\omega_1t_1+\omega_2t_2)}
  \left\langle
    b^{lm}_r (\omega_1)
    b^{\ast lm}_s (\omega_2)
  \right\rangle 
  &=& \int_t^{t+\tau/2}dT\,4\pi\delta(\omega_1-\omega_2) e^{i2\omega T} 
  \left\langle
    b^{lm}_r (\omega_1)
    b^{\ast lm}_s (\omega_2)
  \right\rangle
  \nonumber \\
  &=& 4\pi\delta(\omega_1-\omega_2) e^{i\omega \tau}
  {\sin\omega\tau\over\omega}
  \left\langle
    {\bar b}^{lm}_r (\omega_1)
    {\bar b}^{\ast lm}_s (\omega_2)
  \right\rangle.
  \nonumber \\
\end{eqnarray}
where $\omega\equiv\omega_1=\omega_2$.  In deriving the second
equality, we note that the bombardment must occur between $t$ and
$t+\tau$ and use the shift properties of the Fourier transform to
write $\left\langle b^{lm}_r (\omega_1) b^{\ast lm}_s (\omega_2)
\right\rangle = e^{-2\omega t} \left\langle {\bar b}^{lm}_r (\omega_1)
  {\bar b}^{\ast lm}_s (\omega_2) \right\rangle$ where the transform
${\bar b}(\omega)$ denotes the transform of an event centred about
the temporal origin.  In the limit $\tau\rightarrow0$, this expression
becomes $4\pi\delta(\omega_1-\omega_2)\tau$. Substituting, this back
into equation (\ref{eq:mom1exp}) we can perform one of the $\omega$
integrals straight away.  After rearranging we have
\begin{eqnarray}
  \left\langle \Delta I_j(t+\tau) \right\rangle
  &=&
  -{1\over f_0(\bI)}
  l_j{\bf l}\cdot{\partial f_o\over\partial\bI}
  \left|Y_{ll_2}(\pi/2,0)\right|^2 
  \sum_{\mu\nu}\sum_{rs}
  r^l_{l_2m}(\beta)r^{\ast l}_{l_2m}(\beta)
  W^{l_1\,\mu}_{ll_2m}(\bI) 
  W^{\ast l_1\,\nu}_{ll_2m}(\bI) 
  \times \nonumber \\
  &&
  4\pi\int^{\infty}_{-\infty}d\omega
  {\cal M}^{lm}_{\mu r}(\omega)
  {\cal M}^{\ast lm}_{\nu s}(\omega)
  \left\langle
    b^{lm}_r (\omega_1)
    b^{\ast lm}_s (\omega_2)
  \right\rangle .
  \label{eq:mom1expS}
\end{eqnarray}
The expression for $\left\langle \Delta I_j(t+\tau) \Delta I_k(t+\tau)
\right\rangle$ follows by analogy with the equations
(\ref{eq:mom1exp}) and (\ref{eq:mom2exp}) in \S\ref{sec:pointmass}.

Symmetry suggests the choice of perturbing orbits on the equatorial
plane should not effect the final results.  This can be explicitly
demonstrated explicitly using the rotational properties of the
spherical harmonics.  Let ${\cal
  R}^l_{mm^\prime}(\alpha,\beta,\gamma)$ be the rotation matrix with
the Euler angles $\alpha,\beta,\gamma$.  Then, because
\begin{displaymath}
  Y_{lm}(\theta,\phi) = \sum_{m^\prime=-l}^l 
  Y_{lm^\prime}(\theta^\prime,\phi^\prime) {\cal
  R}^l_{mm^\prime}(\alpha,\beta,\gamma)
\end{displaymath}
where the primed coordinates refer to the rotated coordinate system,
we have
\begin{equation}
  b^{lm}_j(t) = \sum_{m^\prime=-l}^l   b^{lm^\prime}_j(t)
  {\cal R}^l_{mm^\prime}(\alpha,\beta,\gamma)
\end{equation}
For an isotropic spherical system ${\cal M}^{lm}_{\mu r}(\omega)$ is
independent of $m$.  We can exploit this and the rotational properties
of $b^{lm}_r (\omega)$ to simplify the computation of $ \left\langle
  b^{lm}_r (\omega_1) b^{\ast lm}_s (\omega_2) \right\rangle$.  We may
express $b^{lm}_r (\omega)$ in any convenient coordinate system and
use the rotation matrices to rotate this to any orientation:
\begin{equation}
  \left\langle
    b^{lm}_r (\omega_1)
    b^{\ast lm}_s (\omega_2)
  \right\rangle
  = 
  \left\langle
    \sum_{m^\prime=-l}^l R^l_{mm^\prime}(\alpha,\beta,\gamma)
    b^{lm^\prime}_r (\omega_1) 
    \sum_{m^{\prime\prime}=-l}^l R^l_{mm^{\prime\prime}}(\alpha,\beta,\gamma)
    b^{\ast lm^{\prime\prime}}_s (\omega_2)
  \right\rangle
\end{equation}
Summing over all values $m$ for a given $l$ we have
\begin{equation}
  \left\langle
    \sum_{m=-l}^l
    b^{lm}_r (\omega_1)
    b^{\ast lm}_s (\omega_2)
  \right\rangle
  = 
  \left\langle
    \sum_{m^\prime=-l}^l
    b^{lm^\prime}_r (\omega_1) 
    b^{\ast lm^\prime}_s (\omega_2)
  \right\rangle
\end{equation}
having used the orthogonality of rotation matrices.  Finally, we
assume that the ensemble average includes random events from all
directions and therefore only the same event will be correlated in the
computation of $\left\langle b^{lm}_r (\omega_1) b^{\ast lm}_s
  (\omega_2) \right\rangle$.

\section{Summary}
\label{sec:sum}

This paper presents a general equation for noise-driven evolution
which incorporates the full self-gravitating response of a stochastic
process in the perturbation limit.  The motivating question is same as
that of violent relaxation: what is the long-term response of a
stellar system to a fluctuating potential?  By working in the
perturbation limit, the linearity guarantees that the process is
Markovian and therefore the response of the stellar system to repeated
stochastic events is naturally treated using the matrix-method and the
statistic methods for handling general stochastic differential equations.

In general, expansions of integral equations for stochastic processes
yield an infinite number of terms.  A general theorem demonstrates
that the truncation at quadratic order is consistent for a Markov
process and the resulting evolution equation takes the Fokker-Planck
form\footnote{This is also true for two-body collisions {\em if} one
  eliminates strong encounters.}.  Evaluation of the Fokker-Planck
coefficients requires the specifying the response of stellar system
individual events in the stochastic process.

We explicitly develop techniques for two general situations:
quasi-periodic perturbers and transient perturbers.  The former case
is motivated by halo of super-massive black holes (e.g. Lacey \&
Ostriker 1985).  The latter case includes
almost everything else; for example, unbound dwarf encounters
(fly-by), orbiting substructure decaying due to dynamical friction,
disrupting dwarfs, and mixing tidal debris.  There is no practical
constraint on deriving the Fokker-Planck coefficients for transient
noise but that the process can be represented as an expansion in some
biorthogonal basis.  For dwarfs on decaying or unbound orbits, this is
particularly easy and can be done by quadrature.  Alternatively, one
may construct an n-body simulation using the expansion method and let
the simulation produce the time series of coefficients directly.

The companion paper (Paper 2) applies the apparatus described here to
investigate the evolution of haloes during the noisy epoch of galaxy
formation.  We find that both unbound encounters and decaying
substructure drives the halo profile to a self-similar form similar to
those found recently in cosmological simulations.  There are a number
of interesting applications.  For example, the distribution of binary
semi-major axes $a$ in the field star populations is proportional to
$a^{-1}$ over several orders of magnitude.  Preliminary work suggests
that this may be explained as the result of fluctuations from the
noisy environment found in proto-stellar clusters and molecular clouds.
Other possible applications include effect of transient bar formation
and spiral structure on overall galactic structure in the 10 billion
years since formation.

\section*{Acknowledgments}

I thank Enrico Vesperini for comments and discussion.  This work was
support in part by NSF AST-9529328.

\label{lastpage}

\end{document}